\documentclass{osa-article}

\journal{osajournal}

\articletype{Research Article}
\usepackage{color,soul}

\begin{document}

\title{A compact green Ti:Sapphire astro-comb with 43-GHz repetition
frequency}

\author{Eunmi Chae,\authormark{1,2} Eiji Kambe,\authormark{3,4} Kentaro Motohara,\authormark{5,6} Hideyuki Izumiura,\authormark{4} Mamoru Doi,\authormark{5} and Kosuke Yoshioka\authormark{1,*}}

\address{\authormark{1}Photon Science Center, School of Engineering, The University of Tokyo, Bunkyo-ku, Tokyo, 113-8656, Japan\\
\authormark{2}Department of Physics, College of Science, Korea University, Seongbuk-gu, Seoul, 02841, Republic of Korea\\
\authormark{3}Subaru Telescope, National Astronomical Observatory of Japan, 650 North A’ohoku Pl., Hilo, HI, 96720 USA\\
\authormark{4}Okayama Branch, Subaru Telescope, National Astronomical Observatory of Japan, Kamogata, Asakuchi, Okayama 719-0232, Japan\\
\authormark{5}Institute of Astronomy, School of Science, The University of Tokyo, Mitaka, Tokyo 181-0015, Japan\\
\authormark{6}Advanced Technology Center, National Astronomical Observatory of Japan, Mitaka, Tokyo 181-8588, Japan}

\email{\authormark{*}yoshioka@fs.t.u-tokyo.ac.jp} 



\begin{abstract}
A compact green astro-comb with 43-GHz repetition rate is developed based on a Ti:Sapphire optical frequency comb (OFC) and a mode-selecting cavity. 
The OFC’s large repetition rate of 1.6 GHz eases the requirements for the mode-selecting cavity. 
Unnecessary frequency-modes of the OFC are suppressed down to $5 \times 10^{-4}$ at 535 nm -- 550 nm using a single mode-selecting cavity with 70-MHz linewidth. 
The radial velocity precision $\sigma \sim 1.4$ m/s is achieved at the High Dispersion Echelle Spectrosraph for the Okayama 188-cm telescope of the National Astronomical Observatory of Japan using our astro-comb.
With further improvements of the mode-selecting cavity and removal of fiber modal noises, our system will provide a simple, compact, and precise astro-comb setup in visible wavelength region.
\end{abstract}

\section{Introduction}
Since the first discovery in 1995, more than 900 exoplanets have been discovered to date by radial velocity technique using high dispersion $(\lambda/\Delta \lambda > 50,000)$ spectrographs, by which wobble motions of stars caused by planets orbiting them are monitored.
One of the fundamental aims of exoplanet search is to detect earth-mass planets in the so called “Habitable Zone”, which is an orbital region where water can exist in liquid phase.
Such searches for earth-mass planets in Habitable-Zone have been started around M dwarfs \cite{Kotani2018}.
Owing to the low stellar mass and the small radial distance of the Habitable Zone of M dwarfs, the stellar radial-velocity variations caused by such exoplanets reach as high as 1 m/s. 
The newly developed infrared spectrographs are used in these surveys as M dwarfs emit lights largely in infrared wavelength region.

However, the most fascinating but challenging goal of exoplanet search is to detect Habitable-Zone earth-mass planets around solar-like stars, similar to our Sun (a G2V star).
For this purpose, an optical high dispersion spectrograph with a precision of a few tens of centimeters per second or less in radial velocity measurements has to be developed (e.g. 9 cm/s for the Earth around the Sun).
The efficiency of the radial velocity measurements is highest around 400 nm – 500 nm for solar-type stars, considering their peak flux and abundant absorption features that can be used for radial velocity measurements \cite{Bouchy2001}.

To achieve such an extremely high radial velocity precision, optical frequency combs (OFCs) for calibrating spectrometers at observatories, so called astro-combs, have been employed widely since the demonstration \cite{Li2008,Steinmetz2008}.
Many of astro-combs consist of two main components – a fundamental OFC with the repetition frequencies of 100s MHz to several GHz and mode-selecting cavities \cite{McCracken2017}.
Electro-optic frequency combs (EO combs) and micro-cavity OFCs with 10s of GHz repetition frequencies have also shown surprising advances, widely covering infrared wavelength ranges \cite{Suh2019,Metcalf2019}.

Despite the recent advances in various types of astro-combs, construction of an astro-comb at visible wavelengths remains a challenge mainly for two reasons.
The first is that all the fundamental OFCs have spectra in the near- or mid-infrared range so that nonlinear process is inevitable to generate visible wavelengths.
As a result, visible astro-combs developed so far are made from Ti:Sapphire OFCs, amplified fiber OFCs, and EO combs with a nonlinear wavelength conversion such as self-phase modulation and/or second harmonic generation \cite{Metcalf2019, Doerr2012, Probst2016, Glenday2015, McCracken2017a}.
The second reason is that it is very difficult to meet the two requirements for the dispersion of the mode-selecting cavity at visible wavelengths: a sufficient suppression of the unwanted modes of the fundamental OFC and a coverage of wide wavelength range.
The requirements become more challenging when the fundamental OFC has a low repetition frequency ($f_{\textrm{rep}}$).
Typically, multiple mode-selecting cavities are used in-series to overcome the difficulties when a fundamental OFC with a $f_{\textrm{rep}}$ of several hundreds of MHz is employed.
As a result, the overall complexity of the system makes it difficult to maintain stable operation in the harsh environment at the observatory \cite{Doerr2012, Probst2016}.
One can ease the requirements of the mode-selecting cavity by employing an OFC with high $f_{\textrm{rep}}$.
Visible astro-combs based on Ti:Sapphire OFCs with 1-GHz $f_{\textrm{rep}}$ have been demonstrated, achieving either, not both, > 40 dB suppression of unnecessary modes \cite{Glenday2015} or wide wavelength coverage of about 300 nm \cite{McCracken2017a} with fewer numbers of mode-selecting cavities compared to fiber-based astro-combs.

In this manuscript, we report on a proof-of-principle demonstration of an astro-comb consisting of a Ti:Sapphire OFC with a high $f_{\textrm{rep}}$ of 1.6 GHz and a single mode-selecting cavity, implemented at the High Dispersion Echelle Spectrograph (HIDES) for the Okayama 188-cm telescope of the National Astronomical Observatory of Japan (NAOJ) \footnote{The 188-cm telescope was operated by the former Okayama Astrophysical Observatory of the National Astronomical Observatory of Japan until the end of FY2017. Since FY2018 it has been operated under the council of the Tokyo Tech Exoplanet Observation Research Center, Asakuchi City and the National Astronomical Observatory of Japan. It is referred to as "the Okayama 188-cm telescope of NAOJ" or "the 188-cm telescope", hereafter.}.
By employing a Ti:Sapphire OFC with the highest $f_{\textrm{rep}}$ among astro-combs to this day, we were able to build a stable, compact astro-comb with 43 GHz $f_{\textrm{rep}}$ using only a single mode-selecting cavity. 
The spectrum of the astro-comb was observed overnight at HIDES on the 188-cm telescope and utilized to calibrate the spectrometer. 
We will discuss the performance of our proto-type astro-comb with commercial off-the-shelf cavity mirrors and possible improvements for the next generation. 

\section{Experimental setup}

\begin{figure}[htbp]
\centering\includegraphics[width=7cm]{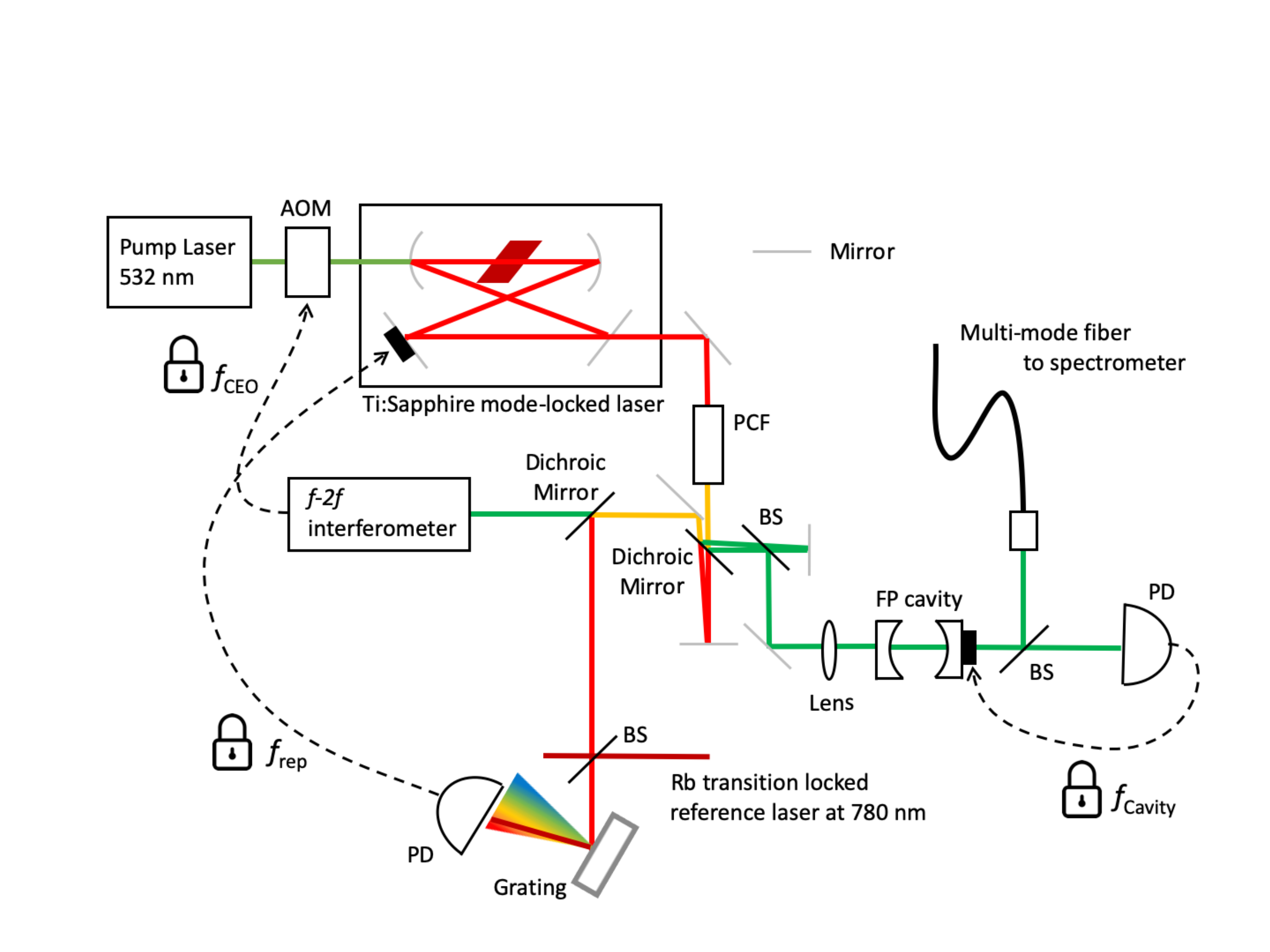}
\centering\includegraphics[width=7cm]{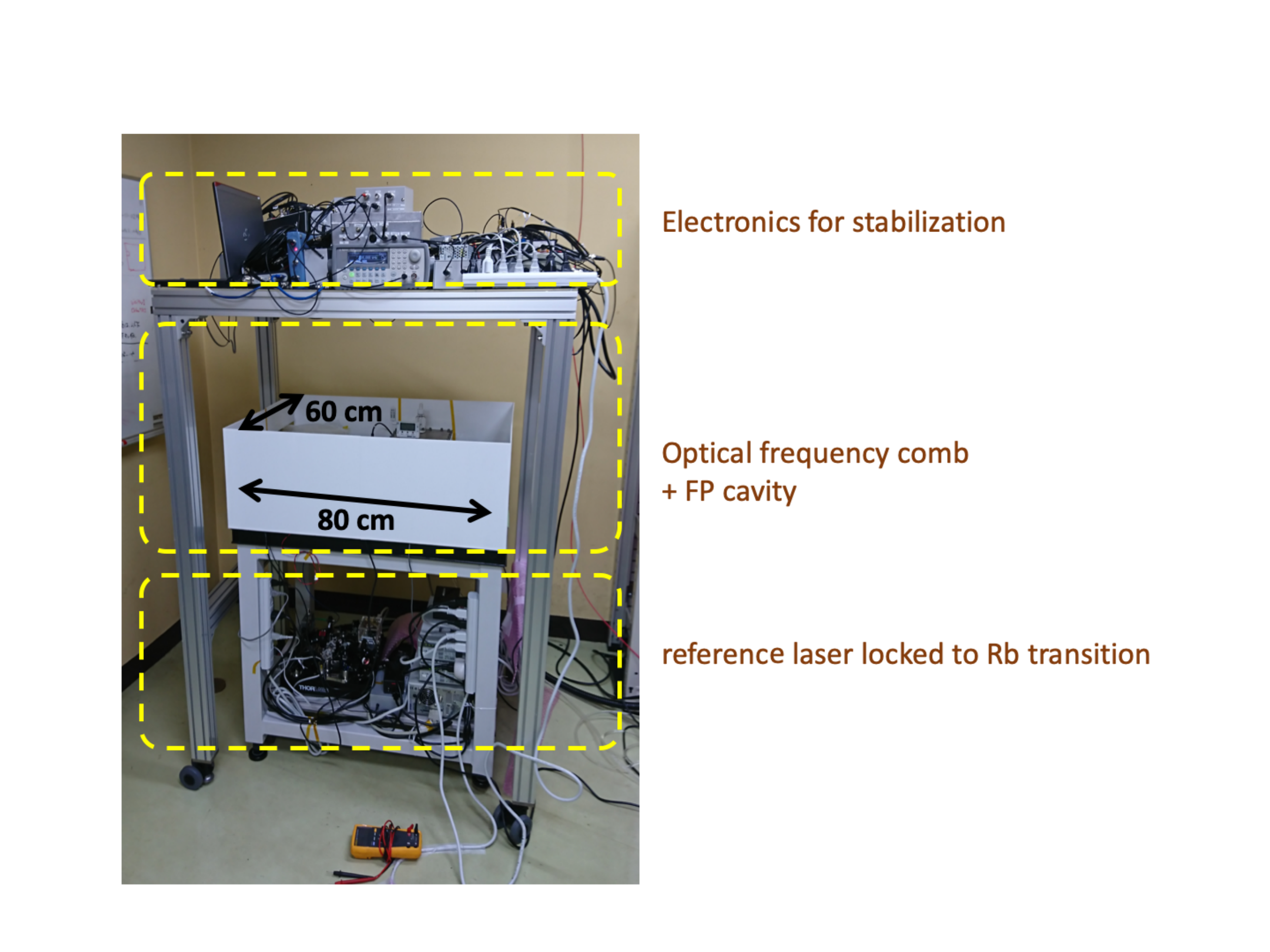}
\caption{Overall experimental setup of the astro-comb. The whole setup is designed to be compact so that it can be fitted within 1 m by 1 m by 2 m (height) space. FP cavity: Fabry-Perot cavity, AOM: acousto-optic modulator, PCF: photonic crystal fiber, BS: beam splitter, PD: photo detector.}
\label{fig:Setup}
\end{figure}

The overall astro-comb setup is depicted in Fig. \ref{fig:Setup}, consisting of a fundamental Ti:Sapphire OFC and a single mode-selecting cavity. 
The details of each component are described in the following sub-sections. 
The whole system is designed to be compact and transportable so that it can be easily shipped and set up promptly at observatories.
The overall setup can be fitted within 1 m by 1 m by 2 m (height) space.
Thanks to this compact and simple setup, the re-adjustment of the astro-comb can be completed within one day after shipping the whole setup to the observatory.

\subsection{Ti:Sapphire-based fundamental optical frequency comb}
\begin{figure}[htbp]
\centering\includegraphics[width=7cm]{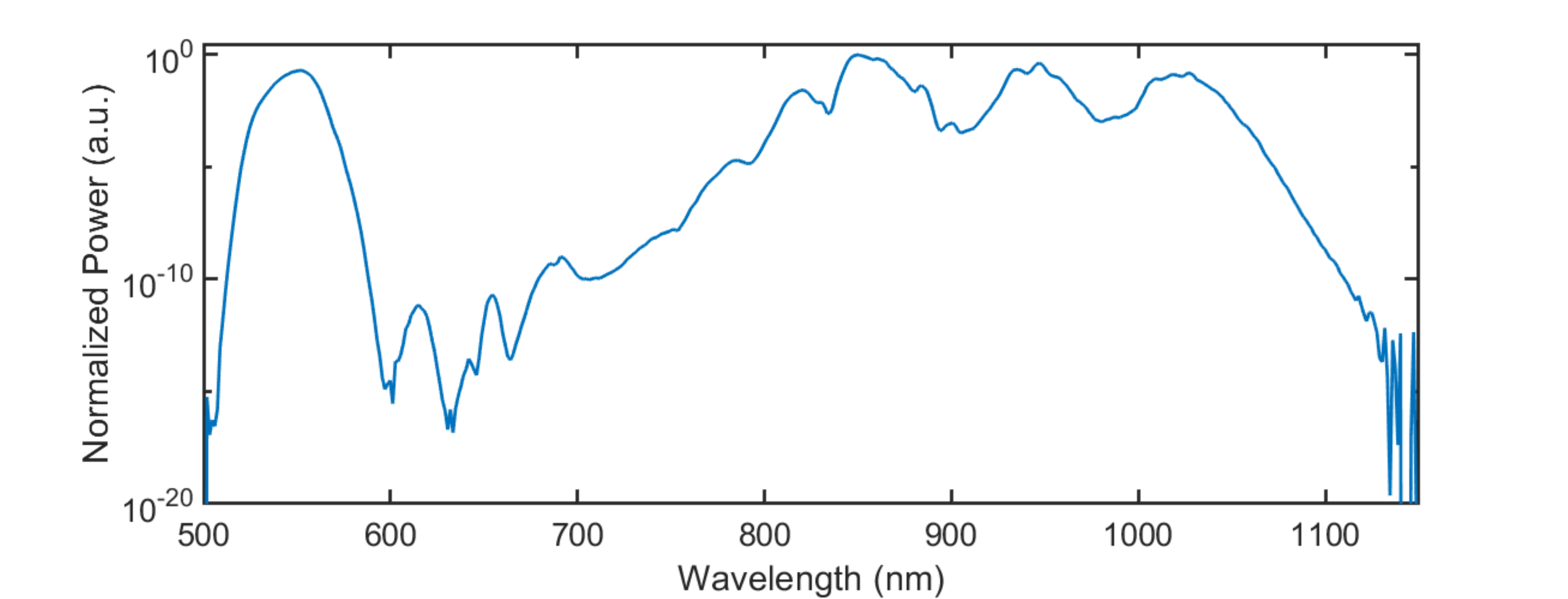}
\caption{Optical spectrum of the fundamental Ti:Sapphire OFC with the $f_\textrm{rep}$ of 1.6 GHz after spectrum broadening by a PCF.}
\label{fig:CombSpec}
\end{figure}

A home-made femtosecond mode-locked Ti:Sapphire oscillator is employed as the fundamental OFC in this experiment.
The round-trip length of the bow-tie laser cavity is about 20 cm, resulting in a high $f_{\textrm{rep}}$ of 1.6 GHz. 
The output pulse has a spectrum that ranges from 700 nm to 900 nm with an output power of about 500 mW at a 6 W pump power at 532 nm (Sprout-G, Lighthouse Photonics).   
The output of the fundamental OFC is frequency-broadened using a photonic crystal fiber (PCF) to cover from 500 nm to 1,100 nm (Fig. \ref{fig:CombSpec}).

The carrier-envelop offset frequency ($f_\textrm{CEO}$) is detected using the self-reference technique.
The $f_\textrm{CEO}$ is stabilized to 327 MHz by applying a feedback signal to an acousto-optic modulator (AOM) that adjusts the pump power.
The remaining degree of freedom of the frequency is stabilized by locking the heterodyne beat between one tooth of the OFC and a continuous-wave (CW) laser at 780 nm stabilized to an atomic transition ($^{87}$Rb 5$^2$S$_{1/2}$ (F=2) $\rightarrow$ 5$^2$P$_{3/2}$ (F=3) at $384.22811520$ THz \cite{Ye1996,Bize1999}) by saturated absorption spectroscopy. 
The heterodyne beat is stabilized at 21.5 MHz. 
As a result, all frequencies of the longitudinal modes of the OFC are precisely determined and stabilized with an $f_\textrm{rep}$ of 1.6174 GHz covering from visible to near IR.
The wavelengths between 530 nm -– 560 nm is used to form the astro-comb in this experiment.

\subsection{Mode-selecting Fabry-Perot Cavity with a 43-GHz free spectral range (FSR)}

\begin{figure}[htbp]
\centering\includegraphics[width=7cm]{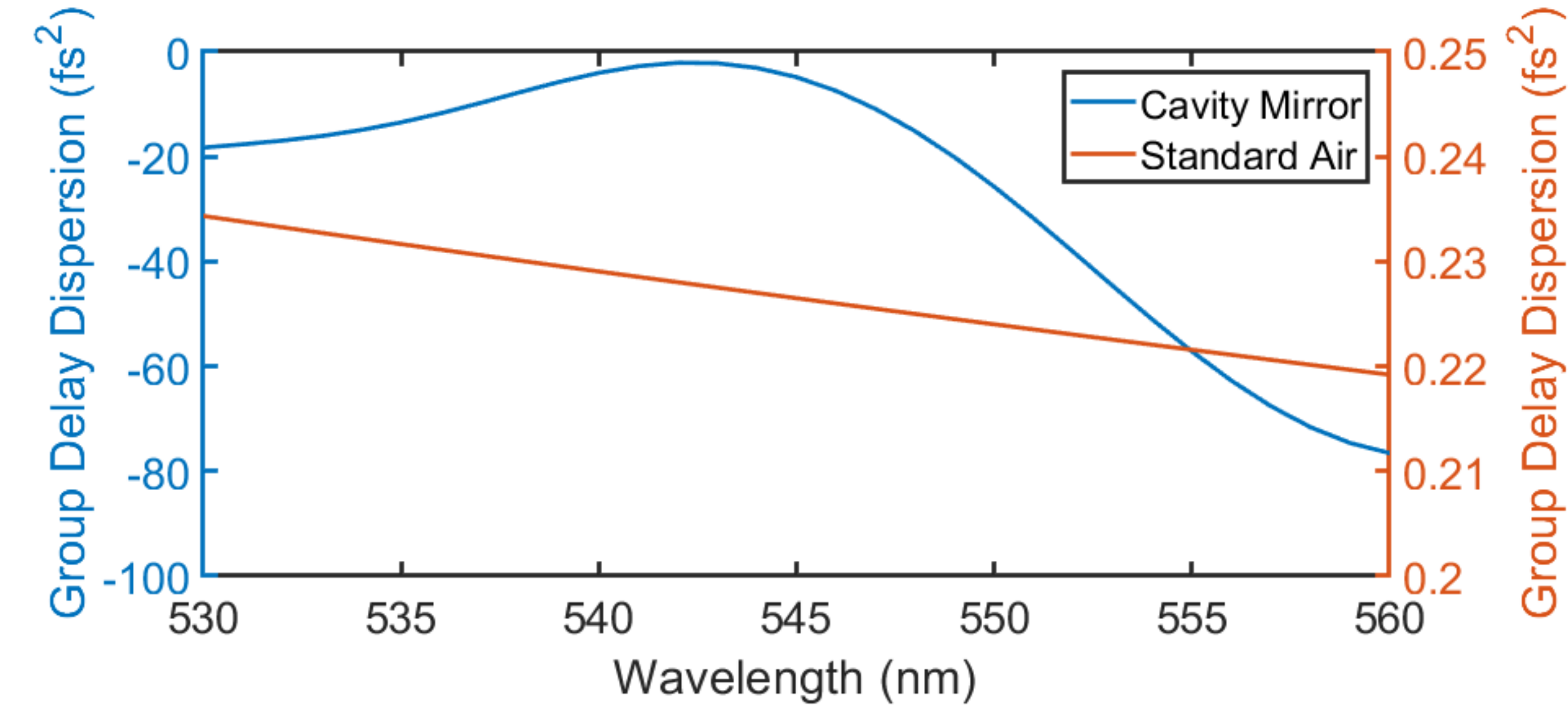}
\centering\includegraphics[width=7cm]{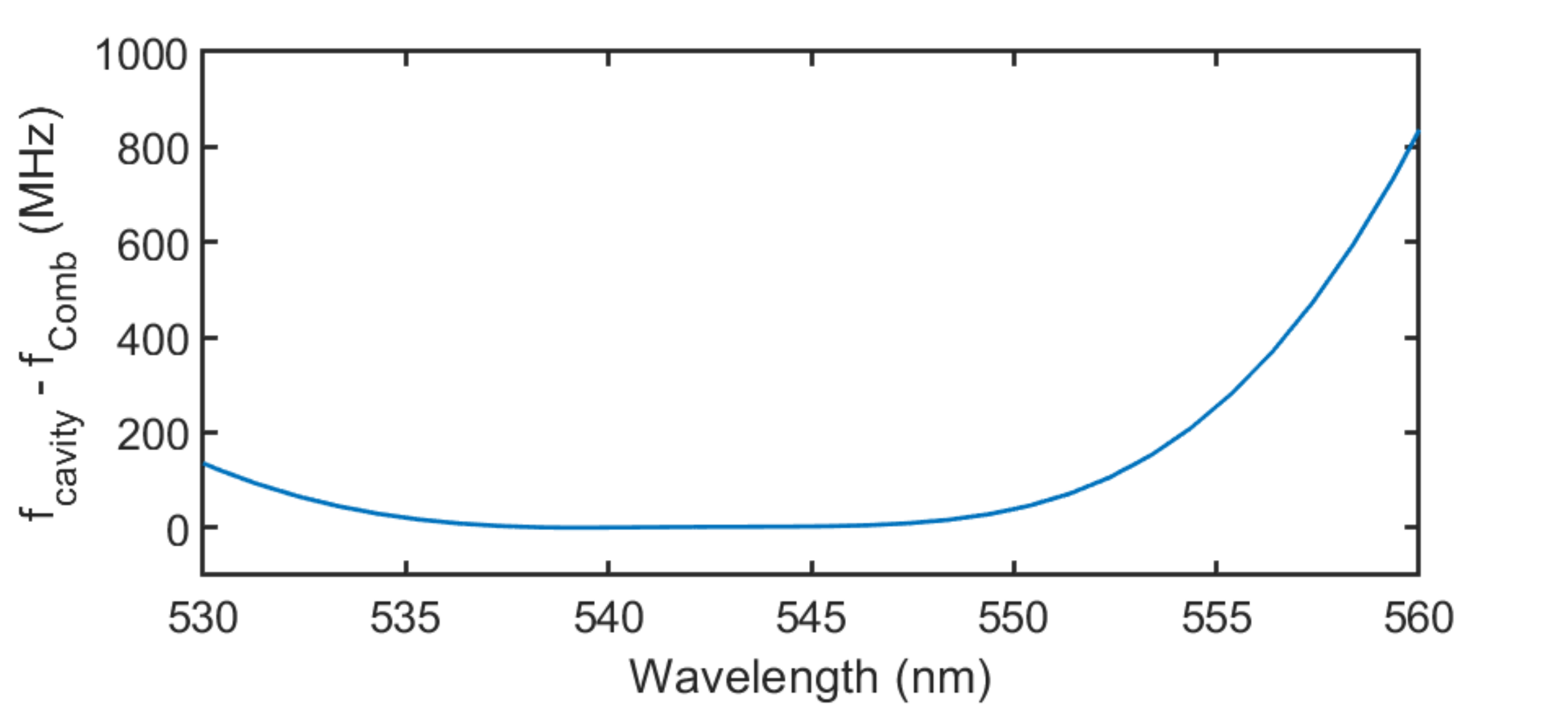}
\centering\includegraphics[width=7cm]{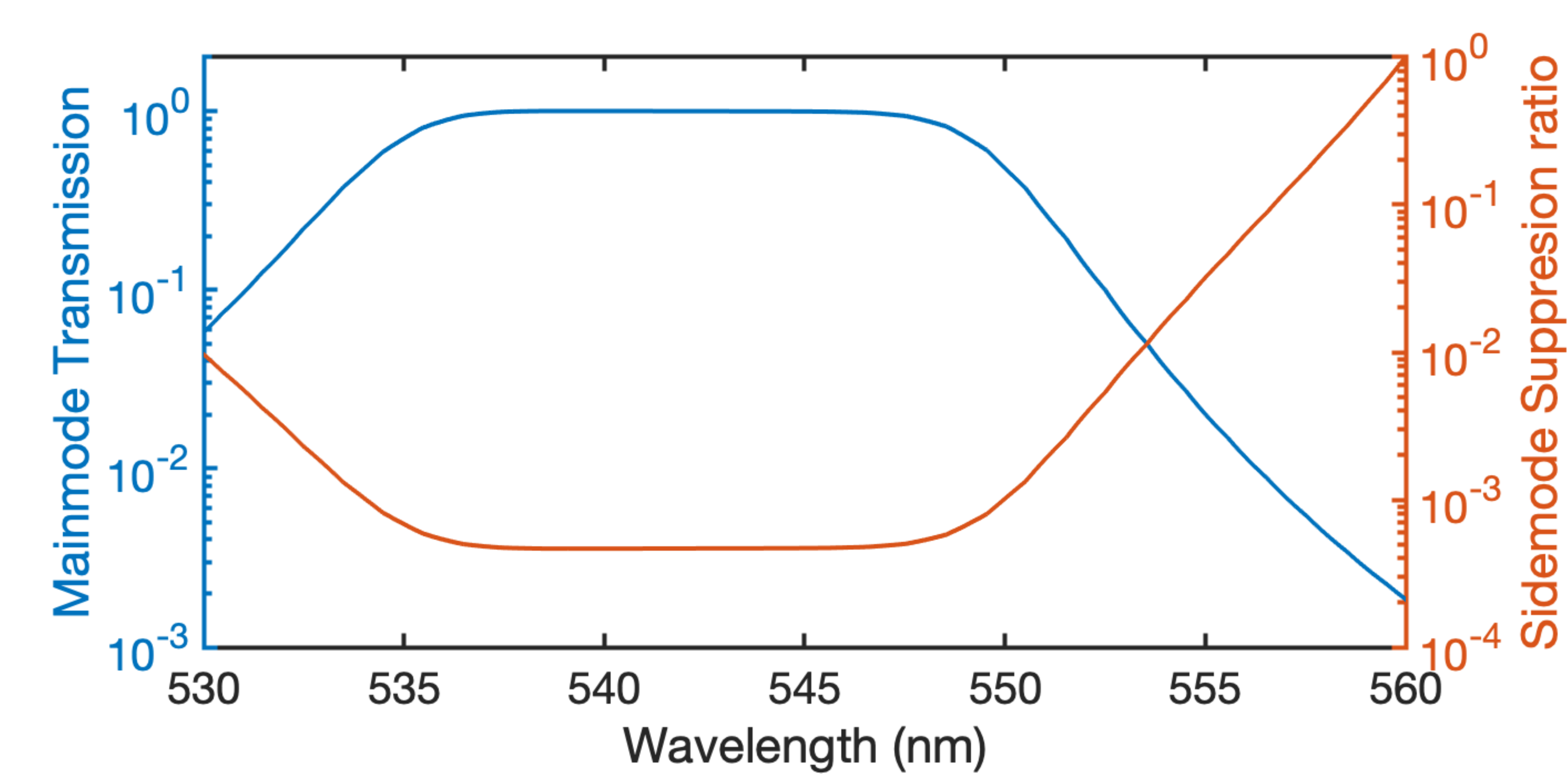}
\caption{(top) Group delay dispersion of the cavity mirrors \cite{Layertec} and the air in between. The values vary from 0 to $-80$ fs$^2$. (middle) Frequency difference between the fundamental OFC and the resonances of the mode-selecting cavity. The difference is larger than the linewidth of the cavity (70 MHz) outside of 535 nm - 550 nm because of the group delay dispersion of the cavity. (bottom) Mainmode transmission and sidemode suppression ratio of the mode-selecting cavity. Sidemode suppression ratio below 10$^-3$ was achieved between 535 nm and 550 nm using the off-the-shelf cavity mirrors with 99.5\% reflectivity thanks to the large $f_\textrm{rep}$ of the fundamental OFC.}
\label{fig:GDD}
\end{figure}

We employed off-the-shelf commercial mirrors (Layertec) to construct the mode-selecting cavity for the astro-comb.
The distance between the two cavity mirrors is set to about 3.5 mm to achieve a free spectral range (FSR) of 43.670 GHz, which is 27 times of the $f_{\textrm{rep}}$ of the fundamental OFC. 
The reflectivity of the cavity mirrors is 99.5\%, corresponding to a linewidth of 70 MHz.
This reflectivity is high enough to suppress the unwanted adjacent modes (sidemode suppression) because of the large $f_{\textrm{rep}}$ of the fundamental OFC.
The total group delay dispersion of the cavity from 530 nm to 560 nm is between 0 and $-80$ fs$^2$ mainly from the cavity mirrors (Fig. \ref{fig:GDD} (top)).
Because of this dispersion, the FSR changes as a function of the wavelength so that it deviates from $n \times$ $f_{\textrm{rep}}$ at some wavelength regions.
This deviation results in discrepancy between the resonant frequencies of the cavity and the frequencies of the OFC (Fig. \ref{fig:GDD} (middle)), causing decrease of the transmission of the OFC as shown in Fig. \ref{fig:GDD} (bottom).
The wavelength dependence of the sidemode suppression ratio is also plotted in Fig. \ref{fig:GDD} (bottom).
One can confirm that the large $f_{\textrm{rep}}$ of the fundamental OFC guarantees adequate main-mode transmissions and sidemode suppressions over several 10s of nm wavelength region even with commercial off-the-shelf cavity mirrors.  

In the experiment, the light of the fundamental OFC with wavelengths shorter than 700 nm is sent to the mode-selecting cavity (Fig. \ref{fig:Setup}). 
A lens is utilized to match the spatial mode of the OFC with that of the cavity. 
The transmitted light from the cavity is split by a 50:50 beam splitter. 
One arm of the beam splitter is used for the stabilization of the cavity length by maximizing the transmitted OFC power using a piezoelectric transducer attached to one of the cavity mirrors. 
The error signal for the stabilization is obtained by dithering the cavity length at 100 kHz. 
The remaining arm of the beam splitter is coupled to a 30-m multimode fiber (FT600EMT, Thorlabs) which sends the light to HIDES.

\subsection{High Dispersion Echelle Spectrograph (HIDES)}

HIDES is an optical, cross-dispersed high dispersion echelle spectrograph of non-white pupil type for the Okayama 188-cm telescope of NAOJ, located in its coude room \cite{Izumiura1999}. 
It was originally fed through the coude optical train of the telescope, and has been so with a fiber-link system that connects the Cassegrain focus of the telescope and the entrance of HIDES \cite{Kambe2013}.
The spectrograph can cover about 400 nm wavelength region (e.g. 360 nm – 760 nm) simultaneously and its maximum wavelength resolution ($\lambda/\Delta \lambda$) is about 110,000 (2 pixel sampling). 
We employ the high efficiency fiber-link (HE mode) to feed stellar lights into HIDES in the experiment. 
As a result, the reciprocal resolution ($\lambda/\Delta \lambda$) of the spectra shown in this paper is about 52,000.
For the experiment, we slightly modified the calibration lamp box so that the astro-comb light and the conventional  wavelength calibration source, Th-Ar lamp, can be switched remotely.

\section{Results and Discussion}

After several test runs at The University of Tokyo, the entire setup was shipped to the Okayama 188-cm telescope of NAOJ. 
Within one day after the arrival, the re-adjustment of the system was completed and the astro-comb was ready for the measurement owing to the simple overall experimental setup. 

The spectra of the astro-comb were recorded every 50 seconds over 84 minutes at HIDES.
The exposure time of each spectrum was one second and it took 49 seconds to read out the data from CCD.
We attempted to take spectra of a bright star and the astro-comb alternately to demonstrate the performance of the astro-comb, but the weather did not permit us to observe the star. 
Hence, we discuss the spectra of the astro-comb alone in this manuscript.
Despite the harsh environment of the observatory with severe vibrations caused by the dome's rotation, the all stabilizations of the astro-comb were stable during the measurements.
The $f_{\textrm{CEO}}$ and the $f_{\textrm{rep}}$ of the fundamental OFC were monitored during the measurement for the determination of the absolute wavelengths of the astro-comb.

\subsection{The spectrum of the astro-comb}

\begin{figure}[htbp]
\centering\includegraphics[width=7cm]{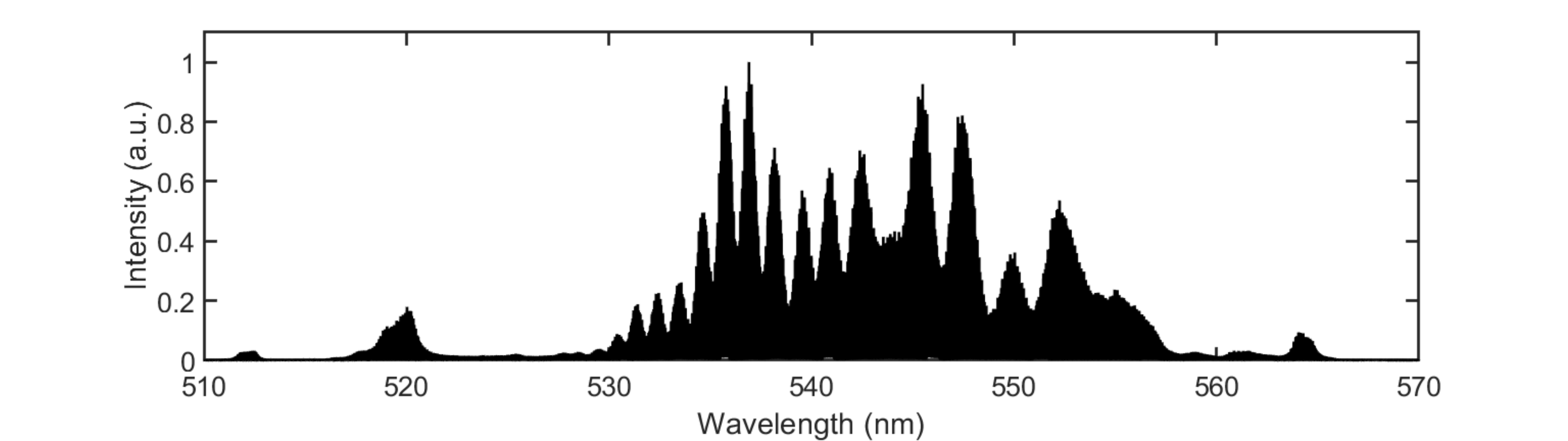}
\caption{Overall spectrum of the astro-comb. The spectrum mainly covers from 530 nm to 560 nm where the fundamental OFC has a high output power.}
\label{fig:AstroCombSpec}
\end{figure}

\begin{figure}[htbp]
\centering\includegraphics[width=7cm]{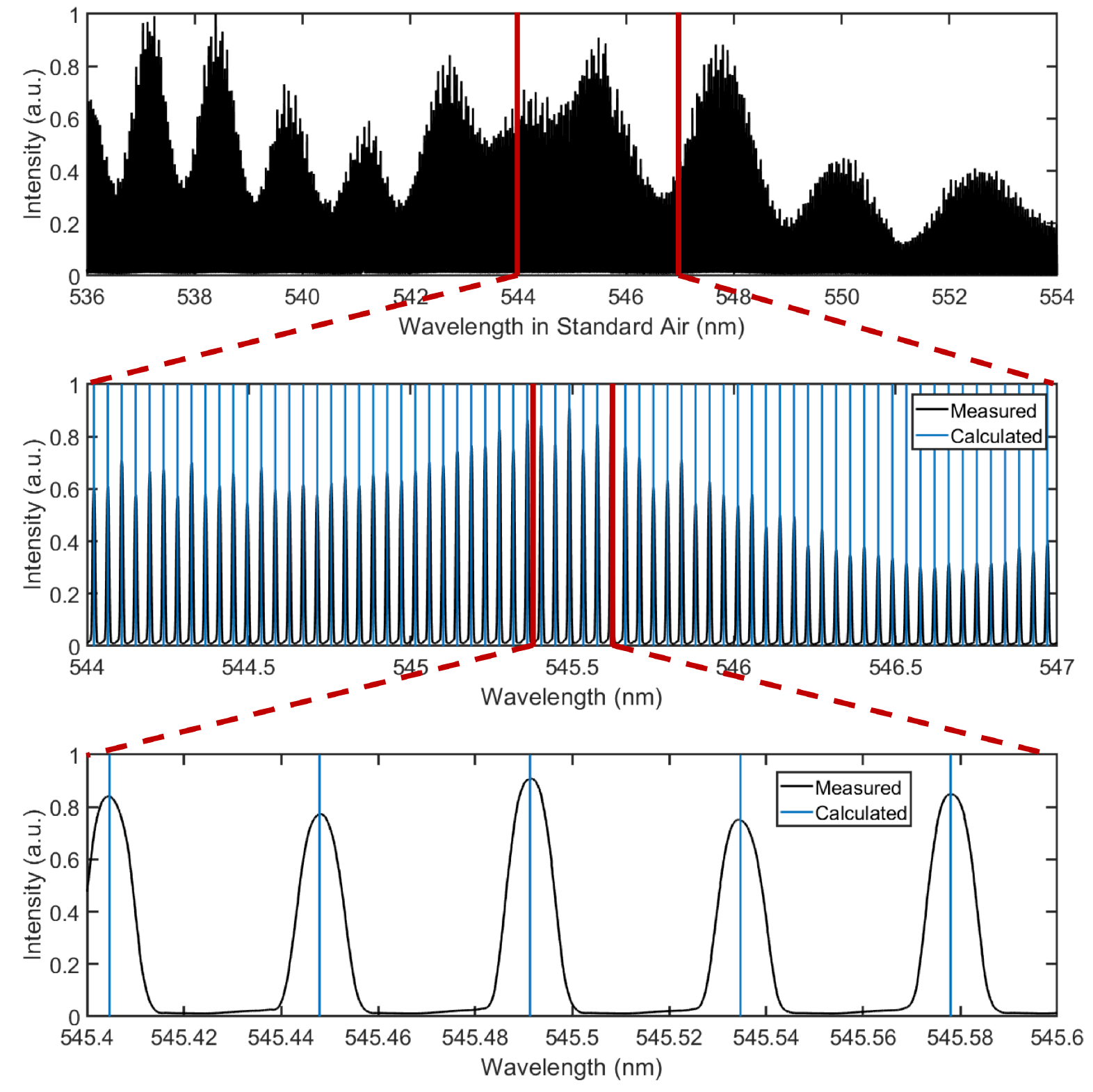}
\caption{Spectra of the astro-comb measured at HIDES. The wavelengths are in standard air. Black lines are the observed spectra by HIDES. Blue lines indicate calculated wavelengths from the parameters of the astro-comb.}
\label{fig:AstroCombSpecZoomed}
\end{figure}

All the data were reduced by the Image Reduction and Analysis Facility (IRAF)\footnote{IRAF is distributed by the National Optical Astronomy Observatories, which are operated by the Association of Universities for Research in Astronomy, Inc., under cooperative agreement with the National Science Foundation.} in a standard manner and wavelengths of the spectra were calibrated, thus the CCD detector pixel positions are mapped to wavelength values, using Th-Ar spectrum obtained just before the sequence of the astro-comb exposures.
Fig. \ref{fig:AstroCombSpec} shows one of the astro-comb spectra obtained by HIDES.
The spectrum spans mainly from 530 nm to 560 nm wavelength range where the fundamental OFC has a high output power. 
When zoomed in, each mode of the astro-comb is clearly resolved with HIDES (Fig. \ref{fig:AstroCombSpecZoomed}).
The observed wavelength separation agrees to the 43.670-GHz frequency interval of the astro-comb which is 27 times of the $f_\textrm{rep}$ of the fundamental OFC. 
The spectrum shows power modulation depending on wavelengths, possibly due to the self-phase modulation of the PCF and the interference of the multimode fiber.

\subsection{Wavelength calibration using calculated frequencies of the astro-comb}

Wavelengths of the all modes of the astro-comb are precisely determined using the parameters of the fundamental OFC and the mode-selecting cavity.
To map the CCD pixel positions to the wavelength using the astro-comb lines, detailed modeling of the observing spectra is necessary.
Here we applied the technique typically used for precise radial velocity measurement using an iodine cell \cite{Butler1996, Kambe2008} in which a stellar spectrum superposed on iodine molecule absorption lines (wavelength reference) is modelled to calculate radial velocity of the star.
We devided the astro-comb spectrum into segments of about 3.6 angstrom width and modelled each segment assuming a Gaussian instrumental profile which broadens the astro-comb lines.
Then, the radial velocity of spectra relative to the first spectrum of the sequence is calculated by averaging the shifts over 51 segments the detailed modelling of which are successfully converged for all the spectra. 
Since the RMS of radial velocity over 51 segments is about 10 m/s, the radial velocity measurement precision of the astro-comb spectrum is about 1.4 m/s in this experiment.
We expect to have the radial velocity precision below 0.5 m/s with 10-fold increase of the available spectral range of the astro-comb. 

The difference between the wavelengths calibrated by the Th-Ar lamp and by the astro-comb is typically about 1/10 pixel, corresponding to $2.6 \times 10^{-4}$ nm in wavelength and 150 m/s in radial velocity. 
This discrepancy is due to the limitation of wavelength calibration with Th-Ar lamp by IRAF which does not take into account the detailed shape of the instrumental profile.

\subsection{Spectrum drift and expected calibration precision }

\begin{figure}[htbp]
\centering\includegraphics[width=7cm]{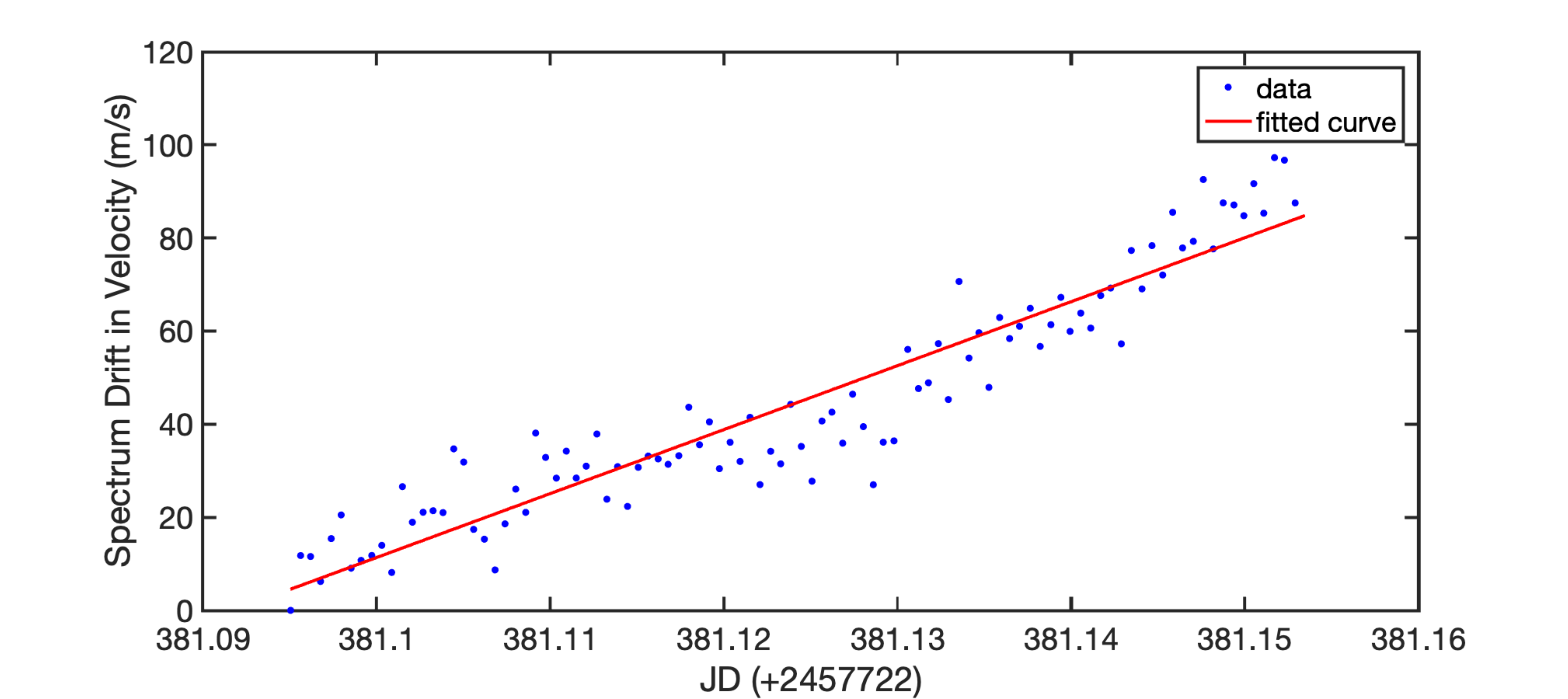}
\caption{Drift of the spectrum at HIDES over 84 minutes. The overall drift indicates the slow change of the spectrometer. The RMS of the residual radial velocity after removing the linear drift is 8 m/s.}
\label{fig:Drift}
\end{figure}

Fig. \ref{fig:Drift} shows the astro-comb spectrum drift on the CCD pixel during the 84-minuite measurement deduced from 3 orders of lines, corresponding to 536 nm - 554 nm.
This wavelength region is selected because of the excellent sidemode suppression ratio within this range.
We also observed a distortion of the line shape profiles outside of this wavelength region in the spectrum measured by HIDES. 
We suspect the distortion is from the fiber-modal noise and the imperfect sidemode suppression due to the dispersion of the cavity mirrors (Fig. \ref{fig:GDD}). 
The overall drift indicates a slow distortion of the spectrograph mainly due to ambient temperature change.
We calculated the RMS of the spectra of the astro-comb after removing the linear drift to estimate the short-term (exposure to exposure) scatter.
The obtained RMS is about 8 m/s, which is smaller than the typical tens of m/s scatter of stellar spectra obtained by HIDES with HE mode.
This scatter is likely due to the fiber modal noise of the multiple mode fiber we used to guide the astro-comb light to the entrance slit of HIDES.
The suppression of this scatter is one of important points to reach a higher precision of the radial velocity measurements.

\subsection{Current limitations and future improvements}

Employing an astro-comb with a wide spectral range is important to improve the precision of the calibration of the spectrometer.
Currently, the spectral range of the astro-comb is limited to about 20 nm because of the uncompensated dispersion of the cavity mirrors. 
This dispersion causes the FSR of the cavity to depend on the wavelengths, leading to the discrepancy of the frequencies between the resonant modes of the cavity and the modes of the fundamental OFC as shown in Fig. \ref{fig:GDD}. 
When the discrepancy exceeds the linewidth of the cavity, the transmission of main modes of fundamental OFC decreases significantly and the sidemode suppression becomes inefficient (Fig. \ref{fig:GDD}). 
The observed spectral range of the astro-comb shown in Fig. \ref{fig:AstroCombSpec} agrees with the range where the discrepancy is below the linewidth of the cavity. 
To cover a wide wavelength range of 500 nm - 700 nm, more careful consideration is required to satisfy the necessary conditions of the mode-selecting cavity. 

There are two requirements for the mode-selecting cavity.
First, the linewidth of the cavity, determined by the reflectivity of the consisting mirrors, should be narrow enough to cut down the unnecessary adjacent modes.
In order to detect the radial velocity change on the order of 10 cm/s, it is required to suppress the adjacent modes below a 10$^{-4}$ level compared to the main mode \cite{Glenday2015}.
Second, the total group delay dispersion of the cavity mirrors and the air in between should be as small as possible, ideally equal to zero, to match the FSR of the cavity and the multiple of the $f_\textrm{rep}$ of the fundamental OFC over the targeted wavelength range. 
In reality, the total dispersion should be tailored so that the main-mode teeth of the OFC locate within the full width at half maximum (FWHM) of the cavity resonance to achieve the sidemode suppression of $10^{-4}$.
As a result, the dispersion requirement becomes stricter when the linewidth of the cavity is narrower. 

When a fiber-based OFC whose $f_\textrm{rep}$s are on the order of 100 MHz, the cavity mirrors with FSR of 40 GHz should have 99.985\% of reflectivity, corresponding to a linewidth of 2 MHz, to achieve the needed $10^{-4}$ sidemode suppression at visible wavelengths.
The total group delay dispersion of the cavity should be tailored within the order of $\pm 0.01$ fs$^2$ over the desired wavelength range in order to match the FSR of the cavity and the $f_{\textrm{rep}}$ of the astro-comb within this narrow linewidth. 
Since engineering the total dispersion at this level is extremely challenging, cascading multiple cavities with lower reflectivity are often employed, which makes the overall setup more complicated \cite{Doerr2012, Probst2016}.

Our Ti:Sapphire OFC with the $f_{\textrm{rep}}$ of 1.6 GHz eases this requirement.
The reflectivity of 99.8\%, which corresponds to a linewidth of 28 MHz, is enough to suppress the unwanted frequency modes of the fundamental OFC down to $10^{-4}$ using a single 40-GHz FSR cavity. 
The required total dispersion of the cavity mirrors is $\pm 0.1$ fs$^2$ over 500 – 700 nm. 

\begin{figure}[htbp]
\centering\includegraphics[width=7cm]{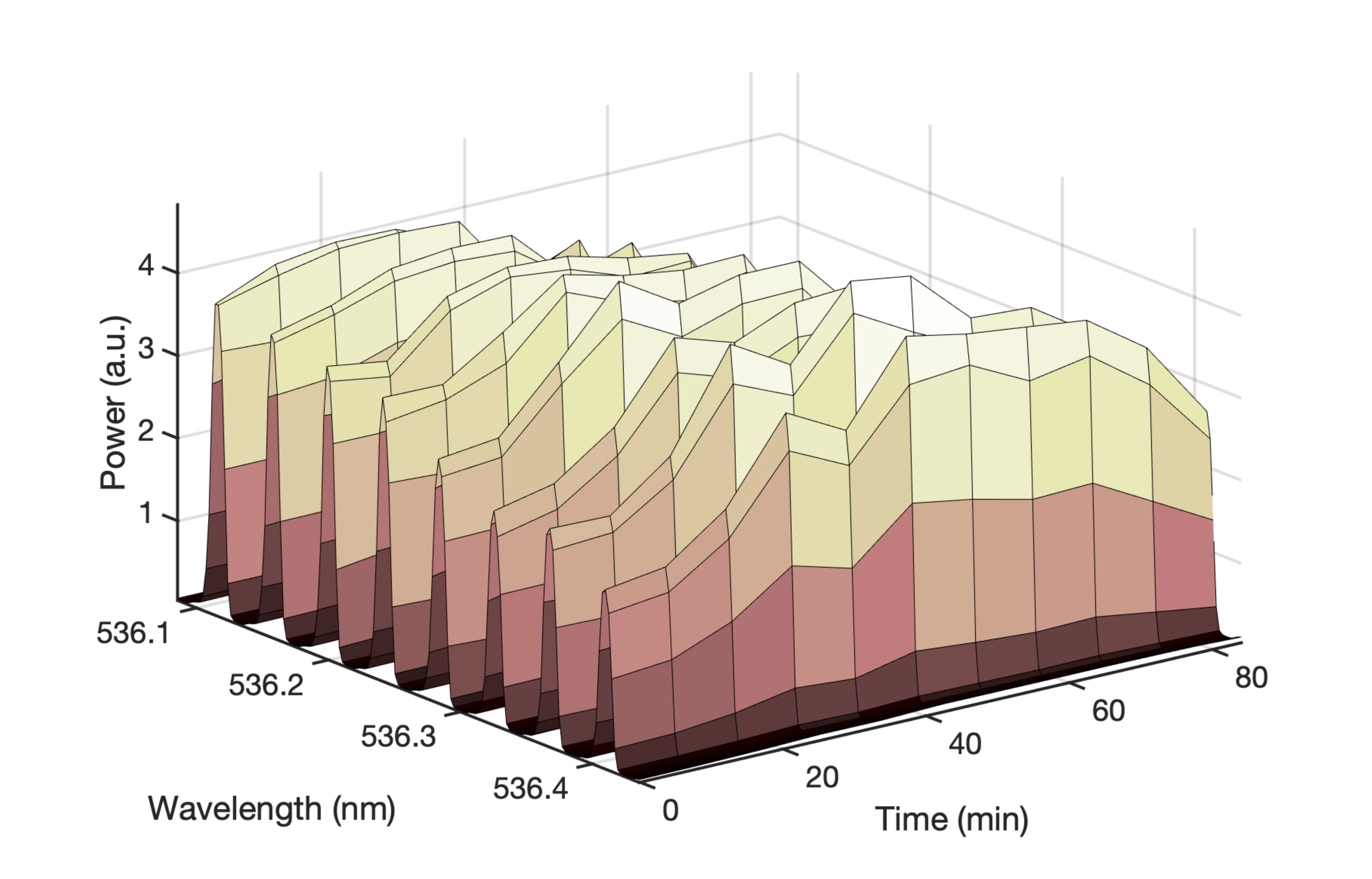}
\caption{Temporal power fluctuations of individual modes of the astro-comb during the measurement. Data taken at every 8.4 minutes during the 84 minutes measurement are used for plotting. }
\label{fig:TemporalChange}
\end{figure}

To reduce the systematic error of the radial velocity, it is desired for each mode of the astro-comb to maintain a stable power.
Fig. \ref{fig:TemporalChange} shows the current temporal power fluctuations of the astro-comb modes during the 84-min measurement.
The power of individual modes measured at HIDES changed gradually even though the stabilization of the cavity remained stable.
This signal power fluctuation can be caused by spatial modal noises of the multimode fiber because of the coherence of the astro-comb. 
Employing a mode-scrambler will remove the power fluctuations and enhance the stability of the astro-comb \cite{Ye2016}.
Also, flattening the spectrum of the astro-comb using a special light modulator would increase the calibration precision for the spectrometer and the sensitivity of the radial velocity measurements.

\section{Conclusion}
As a conclusion, we built and characterized a simple, green 43.670-GHz astro-comb, composed of a Ti:Sapphire OFC and a single mode-selecting cavity. 
The whole system is designed to be compact, fitted within 1 m $\times$ 1 m $\times$ 2 m (height) volume including the needed electronics.
The astro-comb was implemented at HIDES for the Okayama 188-cm telescope of NAOJ within a day of arrival thanks to its simple overall setup.

By choosing a large 1.6174 GHz-repetition frequency of the fundamental OFC, we are able to ease the requirements – the reflectivity and the dispersion – for the mode-selecting cavity.
With commercial off-the-shelf cavity mirrors with a 99.5\% reflectivity, a $5 \times 10^{-4}$ sidemode suppression is achieved at 535 nm – 550 nm. 
The dispersion of the current cavity mirrors limits the wavelength range of the astro-comb to 20 nm, which can be further broadened by employing custom-made dispersion-free cavity mirrors. 
Calibration of HIDES using the astro-comb results in a radial velocity precision of $\sigma \sim $ 1.4 m/s.
With further improvements of the mode-selecting cavity and removal of fiber modal noises, we expect that our system will provide a simple, compact, and precise astro-comb setup in visible wavelength region.

\section*{Funding}
This work was supported by Japan Society for the Promotion of Science (JSPS) KAKENHI Grant Numbers JP20K05357, JP17H06205, The University of Tokyo Excellent Young Researcher Program,
and MEXT Quantum Leap Flagship Program (JPMXS0118067246).
E.C. acknowledges support by National Research Foundation of Korea Grant
Number \raggedright{2020R1A4A101801511} and 2020R1F1A107416211.
H.I. was supported by JSPS KAKENHI Grant Number 16H02169.
E.K. was partially supported by JSPS KAKENHI Grant Number 16H01106.

\section*{Disclosures}
The authors declare no conflicts of interest.

\section*{Data Availability}
Data underlying the results presented in this paper are not publicly available at this time but may be obtained from the authors upon reasonable request.

\bibliography{References-GreenAstrocomb}

\begin{thebibliography}{10}
\newcommand{\enquote}[1]{``#1''}

\bibitem{Kotani2018}
T.~Kotani, M.~Tamura, H.~Suto, J.~Nishikawa, A.~Ueda, M.~Kuzuhara, M.~Omiya,
  J.~Hashimoto, T.~Kurokawa, T.~Kokubo, T.~Mori, Y.~Tanaka, M.~Konishi,
  T.~Kudo, T.~Hirano, B.~Sato, S.~Jacobson, K.~Hodapp, D.~Hall, M.~Ishizuka,
  H.~Kawahara, Y.~Ikeda, T.~Yamamuro, H.~Ishikawa, K.~Hosokawa, N.~Kusakabe,
  J.-I. Morino, S.~Nishiyama, N.~Jovanovic, W.~Aoki, T.~Usuda, N.~Narita,
  E.~Kokubo, Y.~Hayano, H.~Izumiura, E.~Kambe, M.~Ikoma, Y.~Hori, H.~Genda,
  A.~Fukui, Y.~Fujii, H.~Harakawa, M.~Hayashi, M.~Hidai, M.~Machida, T.~Matsuo,
  T.~Nagata, M.~Ogihara, H.~Takami, N.~Takato, H.~Terada, J.~Kwon, D.~Oh,
  K.~Kashiwagi, T.~Nakajima, H.~Baba, and G.~Olivier, \enquote{{The infrared
  Doppler (IRD) instrument for the Subaru telescope: instrument description and
  commissioning results},} in \emph{Ground-based and Airborne Instrumentation
  for Astronomy VII,}  vol. 10702 H.~Takami, C.~J. Evans, and L.~Simard, eds.
  (SPIE-Intl Soc Optical Eng, 2018), p.~37.

\bibitem{Bouchy2001}
F.~Bouchy, F.~Pepe, and D.~Queloz, \enquote{{Fundamental photon noise limit to
  radial velocity measurements},} {\protect\JournalTitle{Astronomy \&
  Astrophysics}} \textbf{374}, 733--739 (2001).

\bibitem{Li2008}
C.-H. Li, A.~J. Benedick, P.~Fendel, A.~G. Glenday, F.~X. K{\"{a}}rtner, D.~F.
  Phillips, D.~Sasselov, A.~Szentgyorgyi, and R.~L. Walsworth, \enquote{{A
  laser frequency comb that enables radial velocity measurements with a
  precision of 1 cm s-1},} {\protect\JournalTitle{Nature}} \textbf{452},
  610--612 (2008).

\bibitem{Steinmetz2008}
T.~Steinmetz, T.~Wilken, C.~Araujo-Hauck, R.~Holzwarth, T.~W. H{\"{a}}nsch,
  L.~Pasquini, A.~Manescau, S.~D'Odorico, M.~T. Murphy, T.~Kentischer,
  W.~Schmidt, and T.~Udem, \enquote{{Laser frequency combs for astronomical
  observations.}} {\protect\JournalTitle{Science (New York, N.Y.)}}
  \textbf{321}, 1335--7 (2008).

\bibitem{McCracken2017}
R.~A. McCracken, J.~M. Charsley, and D.~T. Reid, \enquote{{A decade of
  astrocombs: recent advances in frequency combs for astronomy [Invited]},}
  {\protect\JournalTitle{Optics Express}} \textbf{25}, 15058 (2017).

\bibitem{Suh2019}
M.-G. Suh, X.~Yi, Y.-H. Lai, S.~Leifer, I.~S. Grudinin, G.~Vasisht, E.~C.
  Martin, M.~P. Fitzgerald, G.~Doppmann, J.~Wang, D.~Mawet, S.~B. Papp, S.~A.
  Diddams, C.~Beichman, and K.~Vahala, \enquote{{Searching for exoplanets using
  a microresonator astrocomb},} {\protect\JournalTitle{Nature Photonics}}
  \textbf{13}, 25--30 (2019).

\bibitem{Metcalf2019}
A.~J. Metcalf, C.~D. Fredrick, R.~C. Terrien, S.~B. Papp, and S.~A. Diddams,
  \enquote{{30 GHz electro-optic frequency comb spanning 300 THz in the near
  infrared and visible},} {\protect\JournalTitle{Optics Letters}} \textbf{44},
  2673 (2019).

\bibitem{Doerr2012}
H.-P. Doerr, T.~J. Kentischer, T.~Steinmetz, R.~A. Probst, M.~Franz,
  R.~Holzwarth, T.~Udem, T.~W. H{\"{a}}nsch, and W.~Schmidt,
  \enquote{{Performance of a laser frequency comb calibration system with a
  high-resolution solar echelle spectrograph},} in \emph{Proc. SPIE, Modern
  Technologies in Space- and Ground-based Telescopes and Instrumentation II,}
  vol. 8450 R.~Navarro, C.~R. Cunningham, and E.~Prieto, eds. (SPIE, 2012), p.
  84501G.

\bibitem{Probst2016}
R.~A. Probst, G.~{Lo Curto}, G.~{\'{A}}vila, A.~Brucalassi, B.~L. {Canto
  Martins}, I.~{de Castro Le{\~{a}}o}, M.~Esposito, J.~I. {Gonz{\'{a}}lez
  Hern{\'{a}}ndez}, F.~Grupp, T.~W. H{\"{a}}nsch, R.~Holzwarth, H.~Kellermann,
  F.~Kerber, O.~Mandel, A.~Manescau, L.~Pasquini, E.~Pozna, R.~Rebolo,
  J.~{Renan de Medeiros}, S.~P. Stark, T.~Steinmetz, A.~{Su{\'{a}}rez
  Mascare{\~{n}}o}, T.~Udem, J.~Urrutia, and Y.~Wu, \enquote{{Relative
  stability of two laser frequency combs for routine operation on HARPS and
  FOCES},} in \emph{SPIE Astronomical Telescopes + Instrumentation,
  Ground-based and Airborne Instrumentation for Astronomy VI,}  vol. 9908 C.~J.
  Evans, L.~Simard, and H.~Takami, eds. (SPIE, 2016), p. 990864.

\bibitem{Glenday2015}
A.~G. Glenday, C.-H. Li, N.~Langellier, G.~Chang, L.-J. Chen, G.~Furesz, A.~A.
  Zibrov, F.~K{\"{a}}rtner, D.~F. Phillips, D.~Sasselov, A.~Szentgyorgyi, and
  R.~L. Walsworth, \enquote{{Operation of a broadband visible-wavelength
  astro-comb with a high-resolution astrophysical spectrograph},}
  {\protect\JournalTitle{Optica}} \textbf{2}, 250 (2015).

\bibitem{McCracken2017a}
R.~A. McCracken, {\'{E}}.~Depagne, R.~B. Kuhn, N.~Erasmus, L.~A. Crause, and
  D.~T. Reid, \enquote{{Wavelength calibration of a high resolution
  spectrograph with a partially stabilized 15-GHz astrocomb from 550 to 890
  nm},} {\protect\JournalTitle{Optics Express}} \textbf{25}, 6450 (2017).

\bibitem{Ye1996}
J.~Ye, S.~Swartz, P.~Jungner, and J.~L. Hall, \enquote{{Hyperfine structure and
  absolute frequency of the $^87$Rb 5P$_{3/2}$ state},}
  {\protect\JournalTitle{Optics Letters}} \textbf{21}, 1280 (1996).

\bibitem{Bize1999}
S.~Bize, Y.~Sortais, M.~S. Santos, C.~Mandache, A.~Clairon, and C.~Salomon,
  \enquote{{High-accuracy measurement of the 87Rb ground-state hyperfine
  splitting in an atomic fountain},} {\protect\JournalTitle{Europhysics
  Letters}} \textbf{45}, 558--564 (1999).

\bibitem{Layertec}
{Measurements from Layertec}, \enquote{{LAYERTEC GmbH website},}
  \url{https://www.layertec.de/de/}.

\bibitem{Izumiura1999}
H.~Izumiura, \enquote{{HIDES: a High Dispersion Echelle Spectrograph (C)},} in
  \emph{Observational Astrophysics in Asia and its Future, 4th East Asian
  Meeting on Astronomy (4th EAMA),}  (1999), p.~77.

\bibitem{Kambe2013}
E.~Kambe, M.~Yoshida, H.~Izumiura, H.~Koyano, S.~Nagayama, Y.~Shimizu,
  N.~Okada, K.~Okita, A.~Sakamoto, B.~Sato, and T.~Yamamuro, \enquote{{A Fiber
  Link between the Okayama 188-cm Telescope and the High-Dispersion
  Spectrograph, HIDES},} {\protect\JournalTitle{Publications of the
  Astronomical Society of Japan}} \textbf{65}, 15 (2013).

\bibitem{Butler1996}
R.~P. Butler, G.~W. Marcy, E.~Williams, C.~McCarthy, P.~Dosanjh, and S.~S.
  Vogt, \enquote{{Attaining Doppler Precision of 3 M s-1},}
  {\protect\JournalTitle{Publications of the Astronomical Society of the
  Pacific}} \textbf{108}, 500 (1996).

\bibitem{Kambe2008}
E.~Kambe, H.~Ando, B.~Sato, H.~Izumiura, T.~Sekii, D.~B. Paulson,
  K.~Yanagisawa, S.~Masuda, H.~Shibahashi, A.~P. Hatzes, M.~Martic, J.-C.
  Lebrun, D.~E. Mkrtichian, L.~L. Kiss, H.~Bruntt, S.~J. O'Toole, and T.~R.
  Bedding, \enquote{{Development of Iodine Cells for Subaru HDS and Okayama
  HIDES. III. An Improvement on the Radial-Velocity Measurement Technique},}
  {\protect\JournalTitle{Publications of the Astronomical Society of Japan}}
  \textbf{60}, 45--53 (2008).

\bibitem{Ye2016}
H.~Ye, J.~Han, Y.~Wu, and D.~Xiao, \enquote{{The fiber noise suppression of
  astro-comb fiber link system for Chinese 2.16m telescope},} in
  \emph{Ground-based and Airborne Instrumentation for Astronomy VI,}  vol. 9908
  C.~J. Evans, L.~Simard, and H.~Takami, eds. (SPIE, 2016), p. 99087E.

\end{thebibliography}

\end{document}